\documentclass[amssymb,twocolumn,eqsecnum,aps]{revtex4}
\usepackage{amsmath}
\begin{document}
\title{The Theory of Electrodynamics in a Linear Dielectric}
\author{Michael E. Crenshaw}
\affiliation{US Army Aviation and Missile Research, Development, and Engineering Center, Redstone Arsenal, AL 35898, USA}
\date{\today}
\begin{abstract}
We adopt the continuum limit of a linear, isotropic, homogeneous,
transparent, dispersion-negligible dielectric of refractive index $n$
and examine the consequences of the effective speed of light in a
stationary dielectric, $c/n$, for D'Alembert's principle and the
Lagrange equations.
The principles of dynamics in the dielectric-filled space are then
applied to the electromagnetic Lagrangian and we derive equations of
motion for the macroscopic fields.
A direct derivation of the total energy--momentum tensor from the field
strength tensor for the electromagnetic field in a dielectric is used to
demonstrate the utility of the new theory by resolving the century-old
Abraham--Minkowski electromagnetic momentum controversy in a way that
preserves the principles of conservation of energy, conservation of
linear momentum, and conservation of angular momentum.
\vskip 0.2 cm
\end{abstract}
\maketitle
\vskip 3.0cm
\par
\section{Introduction}
\par
In the real-world, a material is composed of microscopic particles
embedded in the vacuum.
The characteristics of the material are determined by the types of
particles and the manner of their interactions.
Because an explicit accounting of all of the particles and their
interactions is problematic for most materials, we are usually content
with a continuum description of material effects in terms of macroscopic
parameters that are proportional to the number density of the
microscopic particles in a suitably large volume.
Although the macroscopic treatment does not have the same physical
content of the microscopic theory, it must nevertheless be a
self-contained and self-consistent formalism of the physical processes
that occur in the limited system.
\par
In continuum electrodynamics, electromagnetic fields are analyzed using
an empirical set of equations of motion for the fields, the macroscopic
Maxwell equations, in which a simple dielectric is treated as a region
of space where a macroscopic polarization field exists in response to
the presence of a macroscopic electric field.
Alternatively, we can view a stationary dielectric as a continuous
homogeneous region of space in which light travels at a reduced speed,
$c/n$, compared to the speed of light $c$ in the vacuum.
In this article, we derive a self-contained and self-consistent
theoretical treatment of classical continuum electrodynamics from this
fundamental property of a macroscopic dielectric.
The significance of the new continuum electrodynamics is that the
four-dimensional formulation produces the correct traceless symmetric
total energy--momentum four-tensor \cite{BICB,BICB2,BISPIE,BIJMP} that
embodies, in continuum form, the laws of conservation of energy,
conservation of linear momentum, and conservation of angular momentum.
\par
We proceed as follows:
In section II, we adopt the continuum limit of a stationary linear
dielectric of refractive index $n$ and examine the consequences of the
effective light speed $c/n$ for D'Alembert's principle and the Lagrange
equations.
In Section III, we derive equations of motion for the macroscopic 
fields in a stationary dielectric medium
\begin{subequations}
\label{EQf1.01}
\begin{equation}
\nabla\times{\bf B}
+ \frac{n}{c}\frac{\partial{\bf \Pi}}{\partial t}
= \frac{n{\bf J}}{c}
\label{EQf1.01a}
\end{equation}
\begin{equation}
\nabla\times{\bf \Pi}
- \frac{n}{c}\frac{\partial{\bf B}}{\partial t}
= \frac{\nabla n}{n}\times{\bf \Pi}
\label{EQf1.01b}
\end{equation}
\begin{equation}
\nabla\cdot{\bf B}=0
\label{EQf1.01c}
\end{equation}
\begin{equation}
\nabla \cdot {\bf \Pi} = -\frac{\nabla n}{n}\cdot{\bf \Pi}  -\rho
\label{EQf1.01d}
\end{equation}
\end{subequations}
from the Lagrangian.
Here, ${\bf B}=\nabla\times{\bf A}$ is the magnetic field,
${\bf \Pi}= (n/c)\partial{\bf A}/(\partial t)$
is the conjugate momentum field,
$\rho$ is the total charge density,
${\bf J}$ is the free charge current, and
${\bf A}$ is the vector potential.
\par
There is no question that the classical macroscopic Maxwell equations
\begin{subequations}
\label{EQf1.02}
\begin{equation}
\nabla\times{\bf H} - \frac{1}{c}\frac{\partial{\bf D}}{\partial t}
= \frac{{\bf J}}{c}
\label{EQf1.02a}
\end{equation}
\begin{equation}
\nabla\times{\bf E} + \frac{\partial{\bf B}}{\partial t} = 0
\label{EQf1.02b}
\end{equation}
\begin{equation}
\nabla\cdot{\bf B}=0
\label{EQf1.02c}
\end{equation}
\begin{equation}
\nabla \cdot {\bf D}=\rho
\label{EQf1.02d}
\end{equation}
\end{subequations}
successfully explain the phenomena of classical continuum
electrodynamics, with the notable exception of the century-old
Abraham--Minkowski momentum controversy.
The same record of experimental and theoretical validation largely
applies to the set of macroscopic electrodynamic equations of motion,
Eqs.~(\ref{EQf1.01a})--(\ref{EQf1.01d}).
Apart from the scaling of the free charge current, each of the
equations of motion for the macroscopic fields in continuum
electrodynamics, Eqs.~(\ref{EQf1.01a})--(\ref{EQf1.01d}), is
mathematically equivalent to a corresponding Maxwell-Heaviside equation,
Eqs.~(\ref{EQf1.02a})--(\ref{EQf1.02d}) \cite{BIKins,BIFrias}.
However, the transformations do not comprise a tensor transformation and
the two sets of coupled equations of motion are not equivalent.
In Section IV, we develop the tensor form of continuum electrodynamics.
We construct the field-strength tensor and derive the total
energy--momentum tensor, a result that has been sought for over a 
century \cite{BIAM,BIAM2,BIAM3,BIAM4,BIAM5}.
We discuss the content and role of the total energy--momentum tensor 
in terms of the laws of conservation of energy, conservation of linear
momentum, and conservation of angular momentum in a continuum.
\par
\section{Particle Dynamics in a Dielectric Filled Space}
\par
We consider an arbitrarily large region of space to be filled with a
linear, isotropic, homogeneous, transparent dielectric in a regime
in which dispersion, electrostriction, and magnetostriction are
negligible and, for convenience,  we apply the term simple linear
dielectric to this medium.
In the rest frame of the simple linear medium, the constant
refractive index $n$ is the only property of a linear dielectric
that is significant to the current problem.
Let the rest frame of the dielectric be $S(t,x,y,z)$ with
orthogonal axes $x$, $y$, and $z$.
Then position vectors in $S$ are denoted by ${\bf x}=(x,y,z)$.
If a light pulse is emitted from the origin at time $t=0$, then
\begin{equation}
x^2+y^2+z^2-\left ( \frac{ct}{n} \right )^2=0
\label{EQf2.01}
\end{equation}
describes wavefronts in the $S$ system.
Writing time as a spatial coordinate $\bar x_0=ct/n$,
the four-vector $(\bar x_0,{\bf x})=(ct/n,x,y,z)$ represents the
position of a point as a matter of geometry \cite{BIFinn1}.
Because we are using an effective speed of light in defining our
timelike coordinate $\bar x_0$, the macroscopic theory is not, and
should not be expected to be, Lorentz invariant.
Lorentz invariance is tied to the special theory of relativity and
the microscopic Maxwell equations for fields in a vacuum.
A microscopic theory of a dielectric is always possible and such a
theory will be Lorentz invariant as light travels at speed $c$ between
scattering events.
However, Lorentz invariance is not an intrinsic symmetry of a continuous
medium in which the electromagnetic field has been averaged over
multiple scattering events creating a macroscopic field that travels
with an effective speed that is less
than $c$ \cite{BIFinn1,BIFinnEMTensors,BIRosen}.
\par
For a system of particles, the transformation of the position
vector ${\bf x}_i$ of the $i^{th}$ particle to $J$ independent
generalized coordinates is
\begin{equation}
{\bf x}_i={\bf x}_i(\tau;q_1,q_2, \ldots, q_J) ,
\label{EQf2.02}
\end{equation}
where $\tau=t/n$.
Applying the chain rule, we obtain the virtual displacement
\begin{equation}
\delta{\bf x}_i=\sum_{j=1}^J
\frac{\partial {\bf x}_i}{\partial q_j}\delta q_j
\label{EQf2.03}
\end{equation}
and the velocity
\begin{equation}
{\bf u}_i=\frac{d{\bf x}_i}{d\tau}=
\sum_{j=1}^J
\frac{\partial {\bf x}_i}{\partial q_j}
\frac{d q_j}{d \tau}
+ \frac{\partial {\bf x}_i}{\partial \tau}
\label{EQf2.04}
\end{equation}
of the $i^{th}$ particle in the new coordinate system.
Substitution of
\begin{equation}
\frac{\partial{\bf u}_i}{\partial(d q_j/d\tau)}=
\frac{\partial {\bf x}_i}{\partial q_j}
\label{EQf2.05}
\end{equation}
into the identity
\begin{equation}
\frac{d}{d\tau}\left ( m{\bf u}_i\cdot
\frac{\partial{\bf x}_i}{\partial q_j} \right ) =
m\frac{d{\bf u}_i}{d \tau}\cdot
\frac{\partial {\bf x}_i}{\partial q_j}
+
m{\bf u}_i\cdot\frac{d}{d\tau}
\left ( \frac{\partial{\bf x}_i}{\partial q_j}\right ) 
\label{EQf2.06}
\end{equation}
yields
\begin{equation}
\frac{d{\bf p}_i}{d\tau}\cdot
\frac{\partial{\bf x}_i}{\partial q_j} =
\frac{d}{d\tau}
\left ( \frac{\partial}{\partial(d q_j/d \tau)}
\frac{1}{2}m{\bf u}_i^2
\right ) -
\frac{\partial}{\partial q_j}\left ( \frac{1}{2}m{\bf u}_i^2\right ).
\label{EQf2.07}
\end{equation}
\par
For a system of particles in equilibrium, the virtual work of the
applied forces ${\bf f}_i$ vanishes and the virtual work on each
particle vanishes leading to the principle of virtual work
\begin{equation}
\sum_i{\bf f}_i\cdot \delta{\bf x}_i=0
\label{EQf2.08}
\end{equation}
and D'Alembert's principle
\begin{equation}
\sum_i\left ( {\bf f}_i -\frac{d{\bf p}_i}{d\tau}\right )
\cdot \delta{\bf x}_i=0.
\label{EQf2.09}
\end{equation}
Using Eqs.\ (\ref{EQf2.03}) and (\ref{EQf2.07})
and the kinetic energy of the $i^{th}$
particle 
\begin{equation}
T_i= \frac{1}{2} m{\bf u}_i^2,
\label{EQf2.10}
\end{equation}
we can write D'Alembert's principle, Eq.\ (\ref{EQf2.09}), as
\begin{equation}
\sum_j^J \left [ \left ( \frac{d}{d\tau}
\left (
\frac{\partial T}{\partial(d q_j/d \tau)}\right ) 
-\frac{\partial T}{\partial q_j}
\right ) -Q_j \right ] \delta q_j =0 
\label{EQf2.11}
\end{equation}
in terms of the generalized forces
\begin{equation}
Q_j=\sum_i{\bf f}_i\cdot
\frac{\partial{\bf x}_i}{\partial q_j}.
\label{EQf2.12}
\end{equation}
If the generalized forces come from a generalized scalar
potential function $V$ \cite{BIGold}, then we can write
Lagrange equations of motion
\begin{equation}
\frac{d}{d\tau} \left (
\frac{\partial L}{\partial(\partial q_j/\partial \tau)}\right ) 
- \frac{\partial L}{\partial q_j} =0,
\label{EQf2.13}
\end{equation}
where $L=T-V$ is the Lagrangian.
The canonical momentum is therefore 
\begin{equation}
p_j=\frac{\partial L}{\partial(d q_j/d \tau)}
\label{EQf2.14}
\end{equation}
in a linear medium.
Comparable derivations for the vacuum case appear in, for example,
Goldstein \cite{BIGold} and Marion \cite{BIMar}.
\par
\section{Macroscopic Equations of Motion for Fields in a Dielectric}
\par
We consider a dielectric block illuminated at normal incidence from
the vacuum by a quasimonochromatic electromagnetic pulse in the
plane-wave limit.
The simple dielectric medium is linear, isotropic, homogeneous,
transparent, and dispersionless.
Although dielectrics in the real world are much more complicated than
this model of a simple linear dielectric, theoretical physics encourages
reducing the complexity of the real world and eliminating
non-essential details in order to determine what is truly important.
In particular, temporal dispersion is inconsequential for the
arbitrarily long quasimonochromatic electromagnetic field that
is considered here.
The dielectric block is draped with a gradient-index antireflection
coating and spatial variation of the refractive index is sufficiently
smooth that reflection, and the associated radiation
pressure, can be neglected.
Then in the rest frame of the dielectric block, the refractive index
is a smoothly varying, real, and time-independent function of position
in a large, but finite, region of space.
\par
The field theory \cite{BICT,BIHillMlod} is based on a generalization of
the discrete case in which the dynamics are derived from a Lagrangian
density ${\cal L}$.
The generalization of the Lagrange equation, Eq.~(\ref{EQf2.13}),
for fields in a linear medium is \cite{BICT,BIHillMlod} 
\begin{equation}
\frac{d}{d \bar x_0}\frac{\partial{\cal L}}
{\partial (\partial A_j /\partial \bar x_0)}
=\frac{\partial {\cal L}}{\partial A_j}
-\sum_i\partial_{i}
\frac{\partial{\cal L}}{\partial(\partial_{i} A_j )},
\label{EQb3.01}
\end{equation}
where $\bar x_0=ct/n$ is the time-like coordinate in the material
and $x_1$, $x_2$, and $x_3$ correspond to the
respective $x$, $y$ and $z$ coordinates.
We adopt the typical conventions that Roman indices run from one
to three, Greek indices run from zero to three, and $\partial_i$ 
represents the operator $\partial/\partial x_i$.
We take the Lagrangian density of the electromagnetic field in the
medium to be
\begin{equation}
{\cal L}=
\frac{1}{2}
\left ( \left (
\frac{\partial{\bf A}}{\partial \bar x_0} \right )^2
-(\nabla\times{\bf A})^2 \right )+\frac{n{\bf J}}{c}\cdot{\bf A}.
\label{EQb3.02}
\end{equation}
Evaluating the components of Eqs.~(\ref{EQb3.01}), we have
\begin{equation}
\frac{\partial{\cal L}}
{\partial (\partial A_{j}/\partial \bar x_0)}
=\frac{\partial A_j}{\partial \bar x_0} 
\label{EQb3.03}
\end{equation}
\begin{equation}
\frac{\partial \cal L}{\partial A_j}= \frac{nJ_j}{c}
\label{EQb3.04}
\end{equation}
\begin{equation}
\sum_i\partial_{i}
\frac{\partial{\cal L}}{\partial(\partial_{i} A_{j})}
=[\nabla\times\nabla\times {\bf A}]_j
\label{EQb3.05}
\end{equation}
for the Lagrangian density given in Eq.~(\ref{EQb3.02}).
Substituting the individual terms,
Eqs.~(\ref{EQb3.03})--(\ref{EQb3.05}), into Eq.~(\ref{EQb3.01}), the
Lagrange equations of motion for the
electromagnetic field in a dielectric are the three orthogonal
components of the vector wave equation
\begin{equation}
\nabla\times\nabla\times {\bf A}
+ \frac{\partial^2{\bf A}}{\partial \bar x_0^2}
=\frac{n{\bf J}}{c}.
\label{EQb3.06}
\end{equation}
For fields, the canonical momentum density
\begin{equation}
\Pi_j= \frac{\partial{\cal L}}
{\partial (\partial A_{j}/\partial \bar x_0)}
\label{EQb3.07}
\end{equation}
supplants the discrete canonical momentum defined
in Eq.~(\ref{EQf2.14}).
We can write the second-order equation, Eq.~(\ref{EQb3.06}), as a
set of first-order differential equations.
To that end, we introduce macroscopic field variables
\begin{equation}
{\bf \Pi}=
\frac{\partial{\bf A}}{\partial \bar x_0}
\label{EQb3.08}
\end{equation}
\begin{equation}
{\bf B}=\nabla\times{\bf A}
\label{EQb3.09}.
\end{equation}
Obviously, ${\bf \Pi}$ is the canonical momentum field density whose
components were defined in Eq.~(\ref{EQb3.07}) after making the
substitutions indicated by Eq.~(\ref{EQb3.03}).
Substituting the definition of the canonical momentum field ${\bf \Pi}$,
Eq.~(\ref{EQb3.08}), and the definition of the magnetic field ${\bf B}$,
Eq.~(\ref{EQb3.09}), into  Eq.~(\ref{EQb3.06}),
we obtain a Maxwell--Amp\`ere-like law
\begin{equation}
\nabla\times{\bf B}+\frac{\partial{\bf \Pi}}{\partial \bar x_0}
=\frac{n{\bf J}}{c} \,.
\label{EQb3.10}
\end{equation}
The divergence of ${\bf B}$, Eq.~(\ref{EQb3.09}), and the curl of
${\bf \Pi}$, Eq.~(\ref{EQb3.08}), produce Thompson's Law
\begin{equation}
\nabla\cdot{\bf B}=0
\label{EQb3.11}
\end{equation}
and a Faraday-like law
\begin{equation}
\nabla\times{\bf \Pi}
- \frac{\partial{\bf B}}{\partial \bar x_0}
= \frac{\nabla n}{n}\times{\bf \Pi} \, ,
\label{EQb3.12}
\end{equation}
respectively.
We posit the charge continuity law
\begin{equation}
\frac{\partial \rho_f}{\partial \bar x_0}=
-\nabla\cdot\frac{n{\bf J}}{c}
\label{EQb3.13}
\end{equation}
that corresponds to conservation of free charges with a free charge
density $\rho_f$ in the continuum limit.
(Simply multiplying the vacuum charge continuity law by $n$ results
in a discrepancy between the divergence of the variant Maxwell--Amp\`ere
law, Eq.~(\ref{EQb3.12})
and the temporal derivative of the Gauss-like law, Eq.~(\ref{EQb3.16}).)
The divergence of the variant Maxwell--Amp\`ere Law,
Eq.~(\ref{EQb3.10}),
\begin{equation}
\frac{\partial}{\partial \bar x_0}\nabla\cdot{\bf \Pi}=
-\frac{\nabla n}{n}\cdot\frac{\partial {\bf \Pi}}{\partial \bar x_0}
+\nabla\cdot\frac{n{\bf J}}{c}
\label{EQb3.14}
\end{equation}
is combined with the charge continuity law, Eq.~(\ref{EQb3.13}),
to obtain
\begin{equation}
\frac{\partial}{\partial \bar x_0}\nabla\cdot{\bf \Pi}=
-\frac{\nabla n}{n}\cdot\frac{\partial {\bf \Pi}}{\partial \bar x_0}
-\frac{\partial \rho_f}{\partial \bar x_0}.
\label{EQb3.15}
\end{equation}
Integrating Eq.~(\ref{EQb3.15}) with respect to the temporal coordinate
yields a version of Gauss's law
\begin{equation}
\nabla\cdot{\bf \Pi}=-\frac{\nabla n}{n}\cdot {\bf \Pi}-\rho_f -\rho_b,
\label{EQb3.16}
\end{equation}
where $\rho_b$ is a constant of integration corresponding to a bound
charge density.
This completes the set of first-order equations of motion for the
macroscopic fields, Eqs.~(\ref{EQb3.10})--(\ref{EQb3.12}) and
(\ref{EQb3.16}) that were introduced in Sec. I as
Eqs.~(\ref{EQf1.01a})--(\ref{EQf1.01d}).
\par
\section{Field and Energy--Momentum Tensors}
\par
In the Maxwell--Heaviside formulation of classical continuum
electrodynamics, there are two pairs of fields, $\{{\bf E},{\bf B}\}$
and $\{{\bf D},{\bf H}\}$, and two field-strength tensors.
Here, there is a single pair of fields $\{{\bf \Pi},{\bf B}\}$ and a
single field-strength tensor.
The field-strength tensor,
\begin{equation}
F^{\alpha\beta}=
\left [
\begin{matrix}
 0        &\Pi_x      &\Pi_y      &\Pi_z
\cr
-\Pi_x     &0           &-B_z        &B_y     
\cr
-\Pi_y     &B_z         &0           &-B_x       
\cr
-\Pi_z     &-B_y        &B_x         &0        
\cr
\end{matrix}
\right ],
\label{EQb4.01}
\end{equation}
is obtained in the usual way from
\begin{equation}
F^{\alpha\beta}=\partial^\alpha A^{\beta}-\partial^{\beta} A^\alpha
\label{EQb4.02}
\end{equation}
for homogeneous materials.
\par 
The reduction to a single field-strength tensor and a single pair of
fields results in an elegant simplification of four-dimensional
continuum electrodynamics.
For example, the total energy--momentum tensor is defined in terms of
the field tensor by \cite{BIJack,BILL}
\begin{equation}
T^{\alpha\beta}=
-F^{\alpha\lambda}F_{\lambda}^{\beta}+\frac{1}{4}g^{\alpha\beta}
F_{\lambda\nu}F^{\lambda\nu},
\label{EQb4.03}
\end{equation}
such that \cite{BICB,BICB2,BISPIE,BIJMP} 
\begin{equation}
T^{\alpha\beta}=
\left [
\begin{matrix}
({\bf \Pi}^2+{\bf B}^2)/2  &({\bf B}\times{\bf \Pi})_x 
&({\bf B}\times{\bf \Pi})_y  &({\bf B}\times{\bf \Pi})_z
\cr
({\bf B}\times{\bf \Pi})_x    &W_{11}      &W_{12}      &W_{13}  
\cr
({\bf B}\times{\bf \Pi})_y    &W_{21}      &W_{22}      &W_{23}     
\cr
({\bf B}\times{\bf \Pi})_z    &W_{31}      &W_{32}      &W_{33}   
\cr
\end{matrix}
\right ],
\label{EQb4.04}
\end{equation}
where
\begin{equation}
W_{ij}=-\Pi_j\Pi_k-B_jB_k+\frac{1}{2} (\Pi^2+B^2)\delta_{ij}
\label{EQb4.05}
\end{equation}
is the Maxwell stress tensor and $g^{\alpha\beta}$ is the diagonal
metric tensor with non-zero elements $g^{00}=1$ and $g^{ii}=-1$.
\par
The form of the energy--momentum tensor has been debated for over a 
century \cite{BIAM,BIAM2,BIAM3,BIAM4,BIAM5}.
The best known candidates are the 1908 Minkowski \cite{BIMin}
tensor and the 1909 Abraham \cite{BIAbr} tensor. 
However, neither the Minkoswski momentum nor the Abraham momentum is
conserved.
It has been proven that the Gordon momentum \cite{BIAM,BIGord}
\begin{equation}
G_G=\int_{\sigma} \frac{{\bf B}\times{\bf \Pi}}{c} \;dv,
\label{EQb4.06}
\end{equation}
is conserved in our closed system  consisting of a homogeneous
dielectric illuminated by a quasimonochromatic pulse at normal
incidence through a gradient-index
antireflection coating. \cite{BICB,BICB2,BISPIE,BIJMP}.
The total energy
\begin{equation}
U=\int_{\sigma}\frac{1}{2}\left ( {\bf \Pi}^2+{\bf B}^2 \right ) \, dv
\label{EQb4.07}
\end{equation}
is likewise conserved.
Then, in the absence of sources or sinks, the conserved quantities
\begin{subequations}
\label{EQb4.08}
\begin{equation}
U=\int_{\sigma} T^{00} \, dv
\label{EQb4.08a}
\end{equation}
\vskip 0.01in
\begin{equation}
P^i=\frac{1}{c}\int_{\sigma} T^{i0} \, dv
\label{EQb4.08b}
\end{equation}
\end{subequations}
are temporally invariant \cite{BIJack,BILL}.
\par
The homogeneous tensor continuity equation is descriptive of energy
and momentum conservation in an unimpeded flow.
The four-divergence operator for a system with a position four-vector
$(\bar x_0,x,y,z)$ is \cite{BICB,BICB2,BISPIE,BIJMP} 
\begin{equation}
\bar\partial_{\beta}=\left (
\frac{n}{c}\frac{\partial}{\partial t},
\partial_x,\partial_y,\partial_z
\right ).
\label{EQb4.09}
\end{equation}
Then the electromagnetic continuity equations
\begin{subequations}
\label{EQb4.10}
\begin{equation}
\frac{\partial}{\partial \bar x_0}
\left [\frac{1}{2}({\bf \Pi}^2+{\bf B}^2)\right ]+
\nabla\cdot\left ({\bf B}\times{\bf \Pi}\right )=0
\label{EQb4.10a}
\end{equation}
\begin{equation}
\frac{\partial}{\partial \bar x_0}\left ({\bf B}\times{\bf \Pi}\right )
+\nabla\cdot{\bf W}={\bf 0}
\label{EQb4.10b}
\end{equation}
\end{subequations}
are the components of the homogeneous tensor continuity equation
\begin{equation}
\bar\partial_{\beta}T^{\alpha\beta}=0
\label{EQb4.11}
\end{equation}
applied to the total energy--momentum tensor, Eq.~(\ref{EQb4.04}).
Therefore, for a continuous dielectric without charges or currents,
the laws of conservation of energy and linear momentum,
Eqs.~(\ref{EQb4.10a}) and (\ref{EQb4.10b}), are preserved by the
temporal invariance of $U$, Eq.~(\ref{EQb4.08a}), and $P^i$,
Eq.~(\ref{EQb4.08b}).
Meanwhile, conservation of angular momentum follows from the
symmetry $T^{\alpha\beta}=T^{\beta\alpha}$ \cite{BIJack,BILL} of the
energy--momentum tensor, Eq.~(\ref{EQb4.04}) .
\par
The equations of motion for the macroscopic fields, 
Eqs.~(\ref{EQb3.10})--(\ref{EQb3.12}) and (\ref{EQb3.16}) contain
sources that we now add to our homogeneous energy--momentum formalism.
Recognizing that sources affect the conservation of energy and momentum,
we require the sources to be perturbative.
We form scalar products of a field with the equations of motion for
the fields and combine the results to obtain the energy continuity
equation
\begin{equation}
\frac{\partial}{\partial \bar x_0}
\left [\frac{1}{2}({\bf \Pi}^2+{\bf B}^2)\right ]+
\nabla\cdot\left ({\bf B}\times{\bf \Pi}\right )=
\frac{n{\bf J}}{c}\cdot{\bf \Pi}+\frac{\nabla n}{n}\cdot
\left ({\bf B}\times{\bf \Pi}\right )
\label{EQb4.12}
\end{equation}
and the momentum continuity equation
\begin{equation}
\frac{\partial}{\partial \bar x_0}\left ({\bf B}\times{\bf \Pi}\right )
+\nabla\cdot{\bf W}=
\rho{\bf \Pi}+{\bf B}\times\frac{n{\bf J}}{c}
+{\bf \Pi}^2\frac{\nabla n}{n}.
\label{EQb4.13}
\end{equation}
Then, the inhomogeneous tensor continuity equation is
\begin{equation}
\bar\partial_{\beta}T^{\alpha\beta}= f^{\alpha},
\label{EQb4.14}
\end{equation}
where
\begin{equation}
f^{\alpha}=\Bigg (
\frac{n{\bf J}}{c}\cdot{\bf \Pi}+\frac{\nabla n}{n}\cdot
\left ({\bf B}\times{\bf \Pi}\right ),
\rho{\bf \Pi}+{\bf B}\times \frac{n{\bf J}}{c}
+{\bf \Pi}^2\frac{\nabla n}{n}
\Bigg )
\label{EQb4.15}
\end{equation}
is the generalized force four-vector \cite{BICB2}.
The inhomogeneous electromagnetic continuity equations give a general
indication of the effect of sources, but they must be used cautiously.
For example, the gradient of the refractive index must be sufficiently 
small that reflections can be neglected \cite{BICB2}. 
The presence of a charges and currents moving freely in a continuous
medium has been accepted here as a historical imperative and any forces
associated with the charges and currents should be regarded as
perturbative.
\par
\section{Summary}
\par
We have recast classical continuum electrodynamics into a region of
space in which the speed of light is $c/n$, instead of $c$, and derived
equations of motion for the macroscopic electromagnetic fields from
the electromagnetic Lagrangian density.
The success of a new physical theory is often gauged by its ability to
resolve previously intractable problems.
We presented a one-line derivation of the total energy--momentum tensor 
that demonstrates that the new representation is consistent with the
continuum form of the laws of conservation of total energy and total
momentum.
\par

\end{document}